\documentclass[reprint]{revtex4-2}
\usepackage{mathtools}
\usepackage{amsmath,bm}
\usepackage{amssymb}
\usepackage{subcaption}
\usepackage{setspace}
\usepackage{dcolumn}
\usepackage{color}
\usepackage{physics}
\usepackage[hidelinks]{hyperref}
\usepackage{bm} 
\usepackage{amsmath}
\usepackage{physics}
\usepackage[dvipsnames]{xcolor}
\DeclareMathOperator\arctanh{arctanh}
\usepackage{xcolor}
\usepackage{soul}

\begin{document}

\title{Elastic hyperbolic strip lattices}
\author{Nicholas H. Patino$^1$, Luca Lomazzi$^2$, Luca De Beni$^3$, and Massimo Ruzzene$^{1,3}$}
\affiliation{$^1$ Department of Mechanical Engineering, University of Colorado Boulder, Boulder, Colorado 80309, USA}
\affiliation{$^2$ Dipartimento di Meccanica , Politecnico di Milano, Milano, Italy}
\affiliation{$^3$ Department of Aerospace Engineering Sciences, University of Colorado, Boulder, USA}

\date{February 26, 2025}

\begin{abstract}

We investigate the dynamic properties of elastic lattices defined by tessellations of a hyperbolic strip domain. These strip lattices are generated by a conformal map of tessellations of the hyperbolic disk. Their vibrational modes are organized into three distinct classes: boundary-localized, interior-localized, and global. This mode classification is governed by a computed localization index quantifying the spatial localization of each mode along the strip's width. We show that, like hyperbolic lattices in the disk, hyperbolic lattices in the strip exhibit a dynamic spectrum populated by a majority of localized modes. This finding is supported by numerical studies of the dynamics of a strip lattice whose hyperbolically distributed sites are coupled by structural beams. The integrated density of states computation for boundary, interior, and global modes reveals the predominance of localized modes, and the local density of states allows for the identification of spectral bands dominated by particular mode classes. This analysis informs time domain simulations of the lattice response to dynamic forcing by bandlimited inputs dominated by each mode class. The results illustrate distinctive wave propagation behavior when the excited frequency band is dominated by boundary-localized, interior-localized, or global modes. We confirm these observations via vibrometry experiments in the frequency and time domains. In the frequency domain, the measured response confirms that the spectral neighborhoods of each excitation are indeed populated by the mode class predicted by numerical investigations. We further show that the time-averaged responses are consistent with simulations. Through this work, elastic hyperbolic strips emerge as a new class of lattices with characteristic beam-like, truss-core architectures and novel nodal arrangements. The considered configuration shows promising capabilities to confine and guide elastic waves along varying spatial regions depending on the frequency content of excitation.
\end{abstract}

\maketitle

\section{Introduction}

Lattice structures can be organized into periodic arrangements featuring characteristic symmetries~\cite{tao2016design,wadley2006multifunctional} or disordered into amorphous systems lacking symmetry altogether~\cite{zheng2005dynamic,choukir2023interplay,wejrzanowski2013structure}. Somewhere between these two extremes lie organized, deterministic structures that are not periodic. Such aperiodic lattice structures present attractive architectures for material design due to their compelling dynamics and structural properties.

Aperiodicity can be achieved in several ways. One way is to compromise the regularity and local symmetry of a periodic lattice by introducing point or line defects to obtain systems, whether elastic, acoustic, electronic, or photonic, that are characterized by states localized at defect sites which tend to exist at frequencies within bandgaps ~\cite{wu2004splitting,yao2009propagation,qi2016acoustic,shang2020local,bae2017defect,ye2022topological,yin2018band}. Another way to achieve aperiodicity is through quasiperiodic patterning. This method generates deterministic patterns that exhibit rotational symmetry and long range order but lack translational symmetry~\cite{shechtman1984metallic}. Several investigations reveal that quasicrystals exhibit topologically protected edge states \cite{zhou2019topological,xia2020topological} which, for non-floppy modes, span gaps in fractal vibrational spectra~\cite{xia2020topological,bandres2016topological,pal2019topological,davies2022symmetry}. Fractals themselves are another class of deterministic patterns that can be used to obtain aperiodic lattices. Fractal lattices possess order resulting from scale invariance, or self-similarity ~\cite{benoit1977fractals}, whereby a hierarchy of similar features arises at successive fractional length scales, often termed generations, iterations, or epochs. Their corresponding band structures are characterized by multiple gaps, often on the subwavelength scale, as a function of generation~\cite{zhao2020elastic,man2018space}.

In this paper, we investigate a class of aperiodic lattices based on a curved design space mapped to a strip. Specifically, we leverage the hyperbolic plane, a space hosting a constant negative Gaussian curvature, to systematically and deterministically generate aperiodic lattices. In covering this plane continuously and completely with regular tiles, we define a broad--strictly speaking, infinite--set of unique tessellations. When projected onto a flat, Euclidean domain such as the unit disk, these regular hyperbolic tessellations become irregular, aperiodic arrangements by the Euclidean measure. Coupling the tessellation vertices with structural beams defines elastic hyperbolic lattices~\cite{patino2024hyperbolic,ruzzene2021dynamics}. These beams systematically branch out, densify, and shorten towards the lattice's edge, which suggests that these structural assemblies may have the ability to mitigate fracture propagation, increase strength, and enhance energy absorption towards their boundaries~\cite{blonder2011venation,ronellenfitsch2021optimal,xiao2018additively,lei2020parametric}. A recent study~\cite{patino2024hyperbolic} has shown that hyperbolic lattices defined on the unit disk are characterized by a higher proportion of boundary-localized states than their Euclidean lattice counterparts. 

Seeing as circular hyperbolic lattices lend themselves well to localized vibrations, we here leverage a hyperbolic design space in tandem with a conformal map to generate hyperbolic lattices contained within a beam-like domain with potential engineering relevance: the strip. This model has recently gained attention as the ``band model" of the hyperbolic plane~\cite{hubbard2016teichmuller,seppi2019maximal}, though the transformation used to obtain it has been explored in topics of complex analysis for at least a century, where it has come to be known as the hyperbolic ``strip" ~\cite{lewent1925conformal,warschawski1942conformal,nehari1952conformal,opfer1980conformal,gehring1984nehari,chuaqui2011schwarzian,grosche1988path}.

Elastic lattices in the hyperbolic strip are generated by conformally mapping circular hyperbolic lattices to an infinite strip domain where they are truncated to occupy a rectangular area. Though generally aperiodic, we here consider the case where the strip lattice exhibits translational symmetry along its central axis, a consequence of parameters chosen in the conformal map. In fact, this lattice contains all symmetries of the strip: translation, rotation, glide, and horizontal and vertical reflection. Such a symmetry group is known as the seventh Frieze group~\cite{conway2016symmetries}. Frieze patterns are invariant under one of seven Frieze groups. One key advantage of obtaining a Frieze pattern is the admissibility of Bloch modes as a basis for vibrational states, permitting the definition of a dispersion relation~\cite{moore2024acoustic,maurin2017bloch}. Another is that lattices of discrete resonators with Frieze symmetries exhibit topological band structures with protected gaps~\cite{lux2024topological}. While the Bloch bands are not investigated herein, we find computational advantages in considering a Frieze pattern. Since our strip lattice repeats along the real axis, we are only required to generate enough sites to faithfully represent a single cell. This procedure not only significantly reduces the number of required computations, which grows exponentially in generation along with the lattice site count, but also circumvents the issue of empty sites left by mapping a finite lattice to an infinite domain. Generating the lattice in this manner has also been shown to facilitate the conformal mapping of hyperbolic lattices to a ring, which in another vein of research has been used to emulate effective black hole geometries and study their associated gravitational interactions~\cite{dey2024simulating,chen2023ads}.

Following this introduction, the paper is organized in the following sequence: Section II describes the mathematical mapping employed to generate the sites in elastic hyperbolic strip lattices from circular hyperbolic lattices. Section III describes the numerical investigation of the dynamic spectrum of an elastic hyperbolic strip lattice and introduces the localization index employed to classify modes by their relative displacements. We then numerically simulate the time domain response of the lattice to transient inputs in Section IV, where localized behavior is demonstrated. The localization predictions from numerical investigations are then experimentally corroborated by the measurements described in Section V. Finally, Section VI summarizes the main findings of this study and provides recommendations for future investigations.

\section{Generating strip lattices from tessellations of the disk}

The geometry of an elastic hyperbolic strip lattice is obtained by transforming a tessellation of the Poincar\'e disk model of hyperbolic space~\cite{ratcliffe1994foundations,poincare1882memoire}. The Poincar\'e disk is a model of the infinite hyperbolic plane bounded within the complex open unit disk $\mathbb{D}=\{z=re^{i\theta} \in \mathbb{C} \, : \, |z| <1\}$. It is equipped with the non-Euclidean metric $ds^2 = -\frac{4}{K}\frac{|dz|^2}{(1-|z|^2)^2}$ where $ds$ denotes the line element in the hyperbolic plane with Gaussian curvature $K$, here taken as $-1$. In the limit $|z| \to 1$, the metric diverges; hence, the rim represents a curve of ideal points, or points at infinity. 

\begin{figure}[ht]
    \centering
    \includegraphics{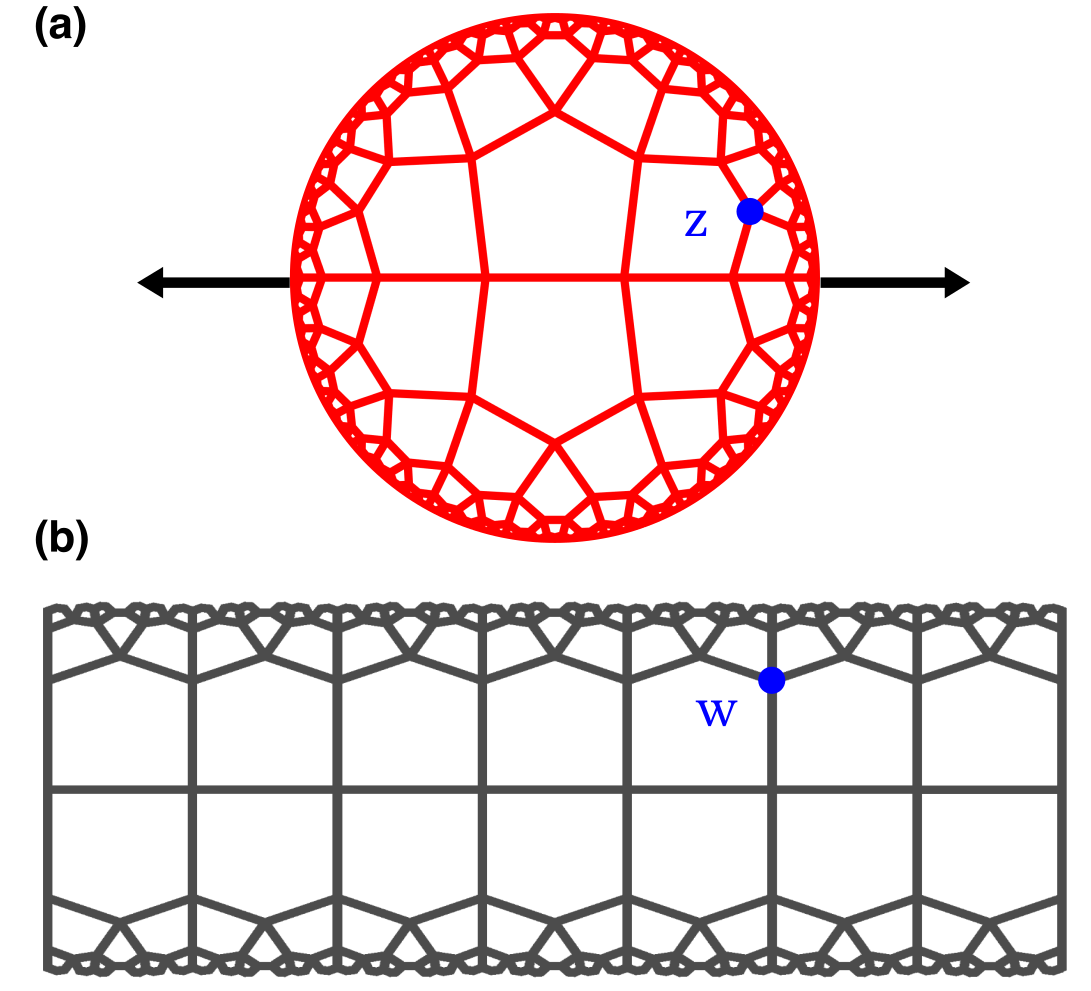} \captionsetup{justification=raggedright,singlelinecheck=false}
    \caption{Mapping of a hyperbolic lattice in (a) the complex unit disk to (b) the complex infinite strip. Points $z \in \mathbb{C}$ in the disk map to points $w\in \mathbb{C}$ in the strip. The two arrows point in the direction of the mapped singularities from $z =\pm$1 to $w=\pm \infty$.}
    \label{Fig1}
\end{figure}

We generate circular hyperbolic lattices~\cite{ruzzene2021dynamics,patino2024hyperbolic} by defining tessellations of the hyperbolic plane, denoted by the Schl\"afli symbol $\{p,q\}$ for $q$ $p$-sided regular hyperbolic polygons meeting at each vertex. Figure~\ref{Fig1}a shows a circular hyperbolic tessellation described by the symbol \{5,4\}. Circular hyperbolic lattices were investigated in two previous studies~\cite{patino2024hyperbolic,ruzzene2021dynamics} where they were shown to exhibit localized wave behavior. This useful property is limited in application as it is restricted to a circular domain. We here seek a mapping to take these lattices from the disk to the strip where their localized wave phenomena are anticipated to persist and whose existence may suggest potential engineering applications in vibration confinement and impact mitigation. 

We map the disk $\mathbb{D}$ to the infinite strip $\mathbb{S}=\{w=x+iy \in \mathbb{C} : |\Im{w}| < 1 \}$ of thickness 2 by the following complex differentiable, or holomorphic, function $w$ of $z$ in $\mathbb{D}$~\cite{balazs1986chaos,mateljevic2020hyperbolic,nehari1952conformal}
\begin{equation}
    w(z) = \frac{4}{\pi} \arctanh{z}.
    \label{arctanh mapping}
\end{equation}

This function contains singularities at $z= \pm1$ which map to $w(\pm 1)= \pm \infty$. The directions normal to the disk at these points can be thought of as ``pulling" directions (illustrated in Fig.~\ref{Fig1}) along which the open disk transforms into an unbounded strip centered about the real axis, aligned with the page width. On the imaginary axis, aligned with the page length, we observe the arresting outcome whereby $\mathbb{D}$ is mapped within the bounds of $-i + \mathbb{R}$ and $i + \mathbb{R}$. A factor of $2/\pi$ ensures that $\pm{i}$ are the nontrivial fixed points of Eq.~\eqref{arctanh mapping} which map to themselves, resulting in the strip's thickness equaling the disk's diameter. This establishes comparable length scales between lattice features in the two domains.

A direct result of Eq.~\eqref{arctanh mapping} being holomorphic with non-vanishing first derivative in $\mathbb{D}$ ~\cite{ahlfors1953complex} is that it is a conformal, or angle-preserving, mapping. Consequentially, the tile and lattice site neighborhoods in $\mathbb{D}$ are preserved in $\mathbb{S}$. As such, the strip is conformally equivalent to the Poincar\'e disk, and it follows that the strip is another model of hyperbolic geometry ~\cite{hubbard2016teichmuller}. In fact, it carries with it the non-Euclidean metric ~\cite{balazs1986chaos,hubbard2016teichmuller,seppi2019maximal,grosche1992energy,dey2024simulating}

\begin{equation}
ds^2 = -\frac{1}{K}(\frac{\pi}{2})^2 \frac{|dw|^2}{\cos^2({\frac{\pi}{2}\Im{w}})},
    \label{strip metric}
\end{equation}

\noindent where again $ds$ is the hyperbolic line, or length, element, and we again take $K=-1$. As in the Poincar\'e disk, the Euclidean distance between equally spaced lattice sites in the hyperbolic plane decreases as one reaches the domain boundary, which in the strip is asymptotic to $|\Im{w}| = 1$. However, along the real axis where $\Im{w}=0$, we find that hyperbolic distances reduce to Euclidean distances scaled by $\pi/2$. This results in the translational symmetry of Fig.~\ref{Fig1}b where the typically distorted hyperbolic lattice constant appears as a constant Euclidean distance between sites on the real axis. As such, translations along the imaginary axis are distorted by the underlying hyperbolic curvature while those along the real axis remain Euclidean translations with a constant scaling.

We define a hyperbolic strip lattice by conformally mapping the lattice sites of a predefined circular hyperbolic lattice to the strip via Eq.~\eqref{arctanh mapping}. The image sites are then connected by straight, Euclidean, distance-dependent couplings. These connections are preserved throughout the mapping by assigning identifiers to the pre-image sites in the disk and storing their connections in an adjacency list. The image sites in the strip carry the same identifiers and hence the same connections. The Euclidean lengths of these couplings vary from those in the disk, but their hyperbolic lengths along geodesics are preserved according to Eq.~\eqref{strip metric}, whereby all coupled sites are equally spaced apart in hyperbolic space. As we realize these lattices in $\mathbb{R}^2$, a gradient-like hierarchy of coupling length scales is observed along the imaginary axis, whereby shorter, stiffer couplings are organized in generations as one approaches the boundary from the real axis (for details on how the masses and stiffnesses of beams depend on generation, see Supplemental Note II). In the circular lattice, a generation is defined as a group of polygons surrounding the nucleation site (the origin), or a previous generation, such that the lattice bulk abides the $\{p,q\}$ description for each site up to those on the boundary~\cite{ruzzene2021dynamics,patino2024hyperbolic}. We maintain this terminology in describing the strip lattice whose generations are defined by iterative groups surrounding the real axis in successive rows. 

Whereas a given lattice in the disk may exhibit rotational symmetry, in the strip it may exhibit translational symmetry along its length, creating a periodic lattice. While this is generally not the case, there are certain pairs of points which we can leverage as singularities of Eq.~\eqref{arctanh mapping} to guarantee a periodic strip. Such pairs are in the set of ideal points lying on symmetry axes of the circular lattice. By purely rotating the circular lattice about its origin, we can align a desired antipodal pair with the singularities $z=\pm1$ (for more details, see Supplemental Note I). In the case of the \{5,4\} hyperbolic lattice, this allows us to generate a periodic \{5,4\} strip lattice, as illustrated in Fig.~\ref{Fig1}. Since the circular lattice is finite, the resulting strip lattice will only repeat up to a certain point along the real axis. To remedy this, we truncate the strip lattice to its unit cell, which is repeated along the real axis six more times. 

\section{Spectral properties and localization of states} 
A numerical model of the considered lattice provides its vibrational spectrum and quantifies the degree of localization and density of states of each eigenmode, or natural vibrational state of the system. The model is developed in the Abaqus environment as a discretized finite element network of 1D Timoshenko beams undergoing out-of-plane deflection, i.e. particle translations perpendicular to the lattice plane. Each beam has a cross-sectional in-plane width of 2.6 mm and an out-of-plane thickness of 3.8 mm. Altogether, the lattice comprises seven unit cells, making it 262.35 mm long and 104.48 mm wide. The material of each beam is modelled as the to-be-used experimental material: Proteus high density polyethylene (HDPE)~\cite{proteus} with Young's modulus $E =$ 1.379 GPa, density $\rho =$ 960 kg/m\textsuperscript{3}, and Poisson's ratio $\nu =$ 0.40. We limit the geometry to a three-generation \{5,4\} strip lattice, as shown in Fig.~\ref{Fig1}b, as the features near the boundary in higher generations become too small to resolve and subsequently manufacture at the considered length scale. At the midpoints of the right and left sides of the lattice, we pin the lattice so as to emulate the mounting conditions of the experimental setup. This is done by prescribing the displacement and moments at those two points such that $u(w =\pm{L/2},t)=0$, $u_{,xx}(w =\pm{L/2},t)=0$, and $u_{,yy}(w =\pm{L/2},t)=0$ where $u(w,t)$ denotes the out-of-plane displacement as a function of position $w $ and time $t$, and $L$ is the lattice length along the real axis. The positions of these essential boundary conditions are in this case purely real as they lie on the central axis of the strip.

Each beam is discretized into a finite element mesh with an element length of 1/2 the shortest lattice beam length, totaling 4115 elements which provides sufficient convergence for the first $N=$ 600 non-rigid body modes. We consider out-of-plane, linear elastic, time-harmonic solutions to the discrete dynamical system of finite elements whose eigenfrequencies $\omega_i=2 \pi f_i$ are given by solving the matrix equation $(\bm{K}-\omega_i^2\bm{M})\pmb{\phi}_i = \bm{0}$ with global stiffness matrix $\bm{K}$ and global mass matrix $\bm{M}$. The eigenmodes are given by $\pmb{\phi}_i$.

\subsection{Localization index and integrated density of states}

To characterize the degree of localization of mode $\pmb{\phi}_i$, we begin by defining the boundary and interior regions of the lattice towards which each mode may localize. Recalling the definition of generations from Section II, we consider $g=3$ generations in the lattice, of which $g\geq2$ comprise the boundary domain $\mathcal{B}$, and $g<2$ comprise the interior $\mathcal{I}$, or central (relative to width), domain. With these regions defined, we apply the following \emph{localization index} to each eigenmode $\pmb{\phi}_i$:

\begin{equation}
   \mathcal{L}_i = \frac{\textstyle\frac{1}{n_{\mathcal{B}}}\displaystyle\sum_{j \in \mathcal{B}}{\abs{\phi^j_i}} - \textstyle \frac{1}{n_{\mathcal{I}}}\displaystyle\sum_{j\in \mathcal{I}}{\abs{\phi^j_i}}}{\frac{1}{n}\displaystyle\sum_{j}{\abs{\phi^j_i}}},
    \label{Localization index}
\end{equation}
where we indicate the $j^{th}$ component of the $i^{th}$ eigenmode by $\phi^j_i$. ${n_{\mathcal{B}}}$ and ${n_{\mathcal{I}}}$ denote the number of boundary and interior nodes, which total all finite element nodes $n = {n_{\mathcal{B}}}+{n_{\mathcal{I}}}$. Eq.~\ref{Localization index} quantifies the degree of localization towards the boundary or interior of the lattice. A positive $\mathcal{L}_i$ indicates a greater mean displacement on the boundary than in the interior, while a negative $\mathcal{L}_i$ indicates a greater mean displacement in the interior than on the boundary. Since $\mathcal{L}_i$ is normalized by the mean displacement of the entire lattice, the range of $\mathcal{L}_i$ lies within $-\frac{n}{n_{\mathcal{I}}}$ and $\frac{n}{n_{\mathcal{B}}}$ which are the two extreme cases with modal displacement exclusively in the interior or boundary respectively. 
In order to classify each state by this localization index, we define a threshold $\mathcal{L}_t = \pm \frac{1}{4}$ which identifies states with a global mean displacement four times larger than that of the boundary or interior. Interior-localized modes are defined by eigenmodes with $\mathcal{L}_i<-\frac{1}{4}$; global modes with $-\frac{1}{4}\leq \mathcal{L}_i \leq \frac{1}{4}$; and, boundary-localized modes with $\frac{1}{4}<\mathcal{L}_i$. The greater the $\mathcal{L}_i$ of a given mode, the more localized it is to its respective domain. 

\begin{figure}
    \centering
    \includegraphics{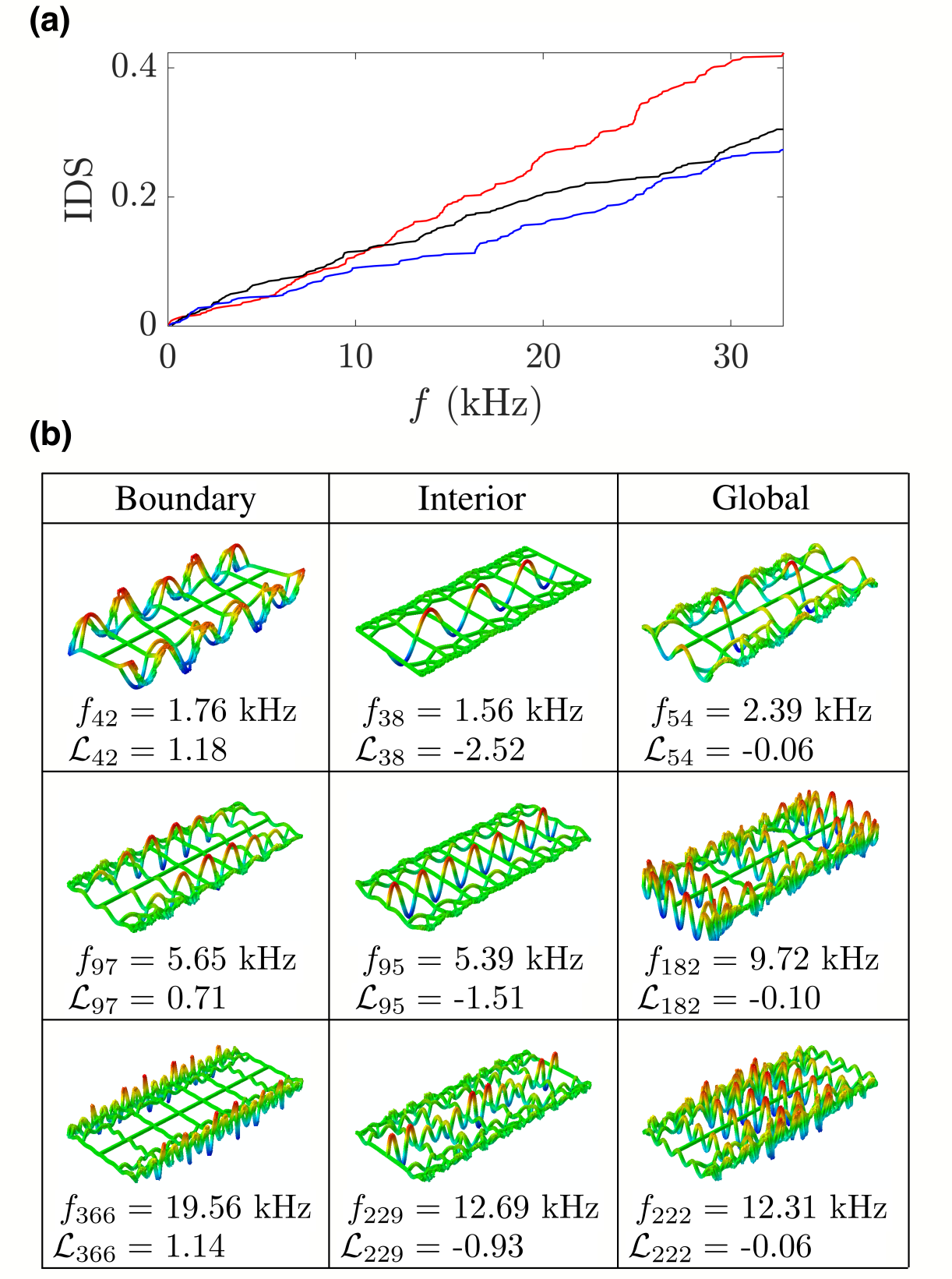}
    \captionsetup{justification=raggedright,singlelinecheck=false}
    \caption{(a) Integrated density of states for boundary modes (red), interior modes (black), and global modes (blue). (b) Examples of boundary, interior, and global modes, normalized by the maximum displacement of each mode (corresponding frequencies and localization indices are provided below each mode shape).}
    \label{Fig2}
\end{figure}

Next, we define an integrated density of states (IDS) \cite{bellissard2001noncommutative,carmona2012spectral,kirsch2007integrated} for each mode class, which is given by

\begin{equation}
    IDS(f) =  \displaystyle\lim_{N\to \infty} \frac{\#\{i \, | \, f_{i} \leq f\}}{N},
    \label{IDS}
\end{equation}

\begin{figure*}
    \centering
    \includegraphics{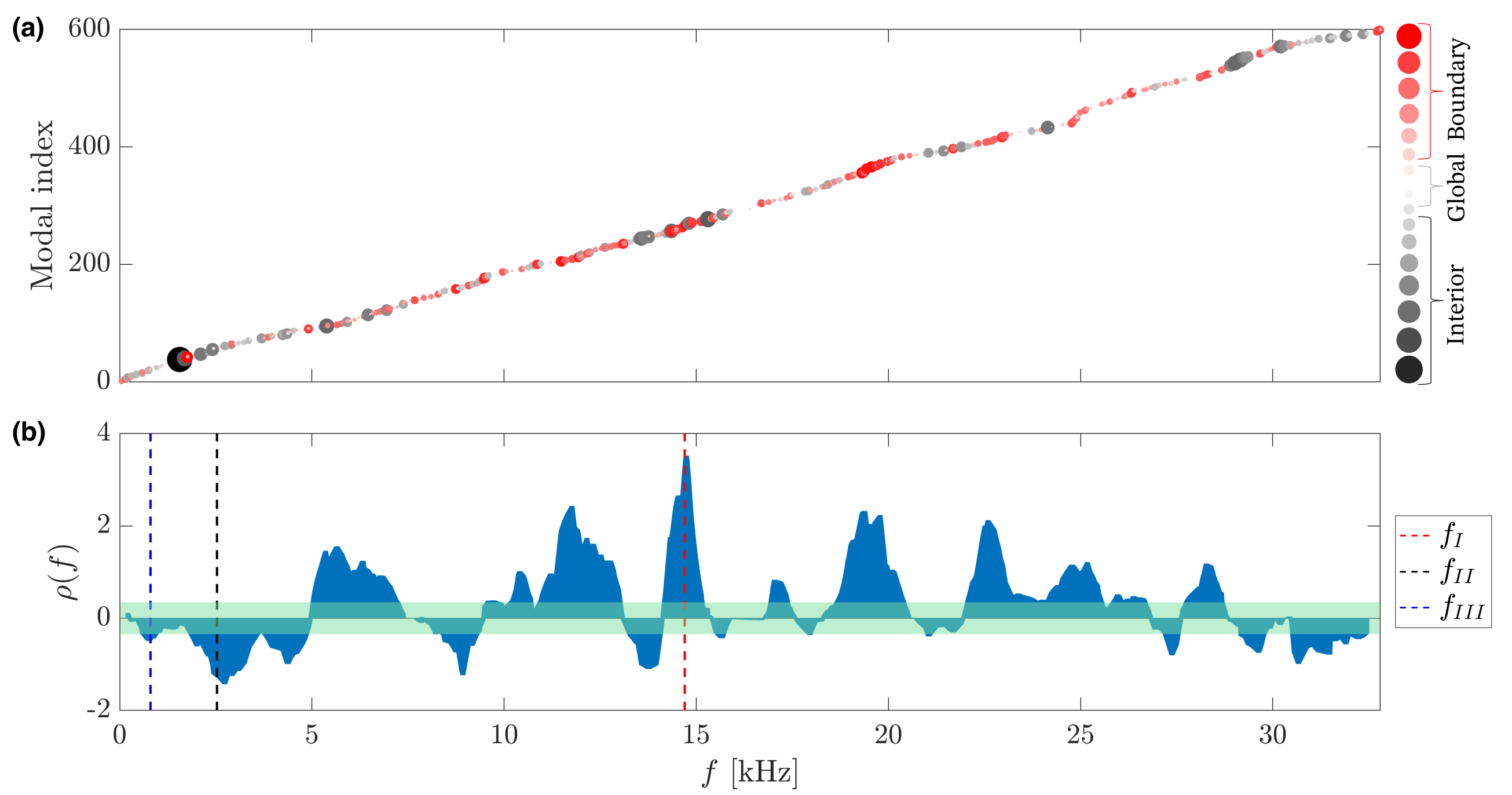}
\captionsetup{justification=raggedright,singlelinecheck=false}
    \caption{(a) Numerical spectrum of the \{5,4\} hyperbolic strip lattice. Eigenmodes are color coded and sized according to their localization index $\mathcal{L}_i$ (red, black, and light pink/white/grey dots respectively denote boundary-localized, interior-localized, and global modes). (b) Relative density of states $\rho(f)$ versus frequency. The global mode regime is highlighted by a transparent mint-colored band. Positive peaks indicate a high density of boundary states relative to interior and global states. Negative peaks indicate a high density of interior states relative to boundary and global states. Three dashed vertical lines indicate central frequencies of inputs applied in numerical simulations and experiments to excite the boundary-dominated region (marked in red at $f_{I}=$ 14.70 kHz), the interior-dominated region (black; $f_{II}=$ 2.49 kHz), and the global-dominated region (blue; $f_{III}=$ 0.80 kHz).}
    \label{Fig3}
\end{figure*}

where $\#$ is a counting operator returning the cardinality $i$ of the set it acts upon. In this case, $\#$ enumerates the eigenfrequencies below frequency $f$. This operation is normalized by the number of considered states $N$, in analogue to a system size, or volume, which is common in discrete resonating systems where the number of degrees of freedom coincides with the number of states. Since a lattice of continuous beams has infinite states, $N$ provides a sufficient approximation of the system size granted that it is large enough~\cite{riva2020adiabatic,pal2019topological,gupta2020dynamics}. By considering $N=$ 600, we sufficiently approximate the IDS profile over the studied frequency range, which converges as $N \rightarrow \infty$. After obtaining the sets of boundary, interior, and global modes via Eq.~\ref{Localization index}, we evaluate their individual IDSs and plot them in Fig.~\ref{Fig2}a. By the 600$^{th}$ mode, around 33 kHz, we observe 73\% localized modes (42\% boundary and 31\% interior), and 27\% global modes. Among the localized modes is a significant class of interior modes which is notably absent from the pre-image disk lattice (see Supplemental Note III). We highlight three interior-localized modes along with three boundary modes and three global modes in Fig.~\ref{Fig2}b, where the corresponding frequencies and localization index values are also reported.

\subsection{Relative density of states}

The numerical spectrum of the considered strip lattice is plotted in Fig.~\ref{Fig3}a where each mode is represented by a dot color-coded and sized by its $\mathcal{L}_i$ value. The color scale is linearly interpolated between red, white, and black, where red is the maximum $\mathcal{L}_i$ (boundary modes), white is $\mathcal{L}_i=0$ (global modes) and black is the minimum $\mathcal{L}_i$ (interior modes). Figure~\ref{Fig3}a provides a useful visualization of the relative degree of localization of different modes, but it does not as easily convey the local density of these modes with respect to frequency. Therefore, in order to convey how spatial localization depends on frequency, we compute the density of states $D(f)$, which provides the number of modes within a specified frequency window. It is given by $D(f) = \mathrm{d}i/\mathrm{d}f$~\cite{kittel1986introduction}, where $i(f)$ is the index of the $i^{th}$ mode. We estimate $D(f)$ numerically by computing a central difference quotient of $i(f)$ given a predetermined frequency window centered at each eigenfrequency~\cite{mulhall2014calculating}. We here choose a window $\Delta f=$ 0.95kHz which is found to be sufficiently small enough to capture changes in density, while not being subject to volatile fluctuations. The resulting numerical formula is thus $D(f)=\frac{\Delta i}{\Delta f}$, with $\Delta i = i(f+\Delta f)-i(f-\Delta f)$. We compute the density of states for each mode class and introduce a non-dimensional measure, which we refer to as the \textit{relative density of states}
\begin{equation}
    \rho(f) = \frac{\Delta i_{\mathcal{\tilde{B}}}(f) -\Delta i_{\mathcal{\tilde{I}}}(f)}{1+\Delta i_{\mathcal{\tilde{G}}}(f)}.
\end{equation}
where we here use the subscripts $\mathcal{\tilde{B}}$, $\mathcal{\tilde{I}}$, and $\mathcal{\tilde{G}}$ to indicate the subsets of boundary, interior, and global modes respectively (adopting tilde notation to differentiate these subscripts from those in Eq.~\eqref{Localization index}). The relative density of states is plotted for the considered lattice in Fig.~\ref{Fig3}b, where positive $\rho(f)$ values indicate a greater local density of boundary modes than interior modes, and negative $\rho(f)$ values indicate a greater local density of interior modes. The magnitude of $\rho(f)$ is not only a function of the difference of boundary and interior-localized state densities but also the global state density, which acts to attract $\rho(f)$ to zero in the case of global state dominance. We introduce a threshold in order to classify frequency regions based on the prevalence of modes in each class. We choose $\rho_t(f)= \pm$ 1/3 which bounds an interval wherein we can expect between 2-3 times more global modes than the difference of boundary and interior modes. The threshold region is highlighted in Fig.~\ref{Fig3}b by a transparent mint band. 

By seeking peaks in $\rho(f)$, we can identify frequency intervals in the spectrum with an affinity for boundary or interior modes. Frequency regions inside or near the threshold region indicate a prevalence of global modes or a balance between boundary and interior modes. In Fig.~\ref{Fig3}b, we mark three frequencies ($f_{I}$, $f_{II}$, and $f_{III}$) in regions of the spectrum which $\rho(f)$ indicates as boundary-dominated, interior-dominated, and global-dominated, plotted as red, black, and blue dashed vertical lines respectively. The red and black lines are selected as they are in the region of the maximum and minimum of $\rho(f)$. The blue line is selected as it is surrounded by a large global-dominated $\rho(f)$ bandwidth and centers frequency content which is spectrally well-separated from the other selections (see Fig. \ref{Fig4}b). These frequencies will later serve as central frequencies for band-limited transient signals in time-domain simulations and experiments demonstrating boundary-localized, interior-localized, and global responses to dynamic inputs.

\section{Numerical results in the time domain}
\begin{figure}
    \centering
    \includegraphics{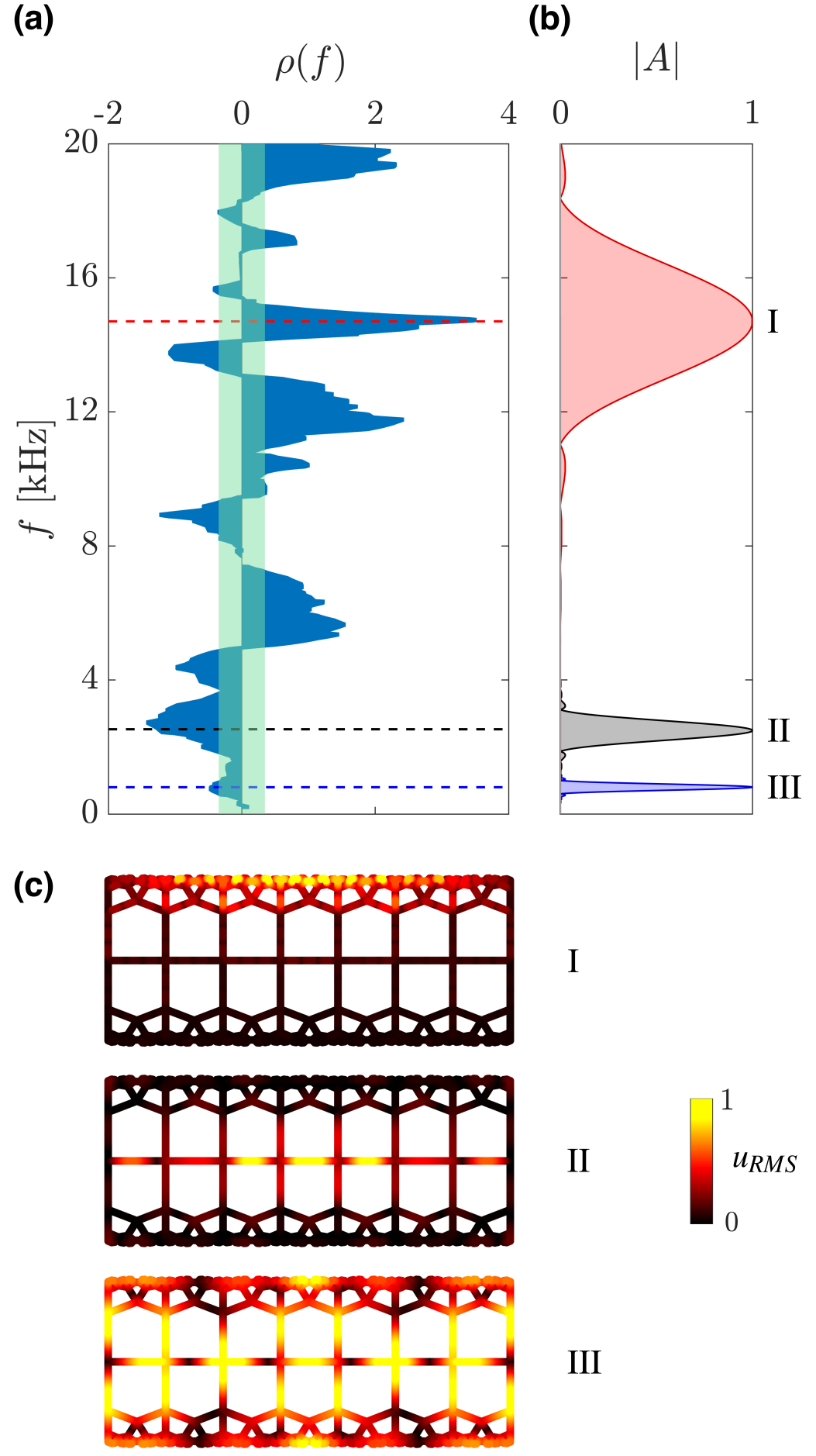}  \captionsetup{justification=raggedright,singlelinecheck=false}
    \caption{(a) Relative density of states in the frequency range of three transient inputs (see Fig.~\ref{Fig3}). (b) Spectral representation of each input signal, labeled I, II, and III and centered at 14.70 kHz, 2.49 kHz and 0.80 kHz respectively. (c) Lattice response to signal I applied to the center of the top boundary, signal II applied to the centroid, and signal III applied to the center of the top boundary. The lattice response color scale represents the normalized RMS displacement of the response throughout the simulated time.}
    \label{Fig4}
\end{figure}

We numerically investigate the time-domain response of the lattice to transient pulse excitations to evaluate the extent to which localized and global modes influence the lattice's dynamic response. We select three frequency regions to excite which are centered in boundary-dominated, interior-dominated, and global bandwidths. The central frequencies are given by the dashed frequency lines in Fig.~\ref{Fig3}b and Fig.~\ref{Fig4}a, which, as noted in Section III, were selected based on $\rho(f)$. Each signal is windowed in the time domain to eight cycles of the central frequency. This in turn gives each signal equivalent fractional bandwidths in the frequency domain. The frequency content of each signal (labeled I, II, and III) is plotted in Fig.~\ref{Fig4}b in the form of a normalized amplitude spectrum. Figure~\ref{Fig4}c shows the root mean square (RMS) displacement field for numerical simulations in the time domain for each signal. Signals I and III are applied as point loads exerted on the center of the top boundary while signal II is a point load applied to the centroid. We note that signal I excites a response confined along the boundary to which it is applied. Signal II leads to a response confined to the interior, with the highest average displacement within the interior of the three central unit cells and the next highest average displacement in the interior of the following pair of cells on either side. Lastly, signal III produces a global response of the lattice. These results confirm the predictions of the relative density of states.

\section{Experimental results}

\begin{figure}
    \centering
    \includegraphics{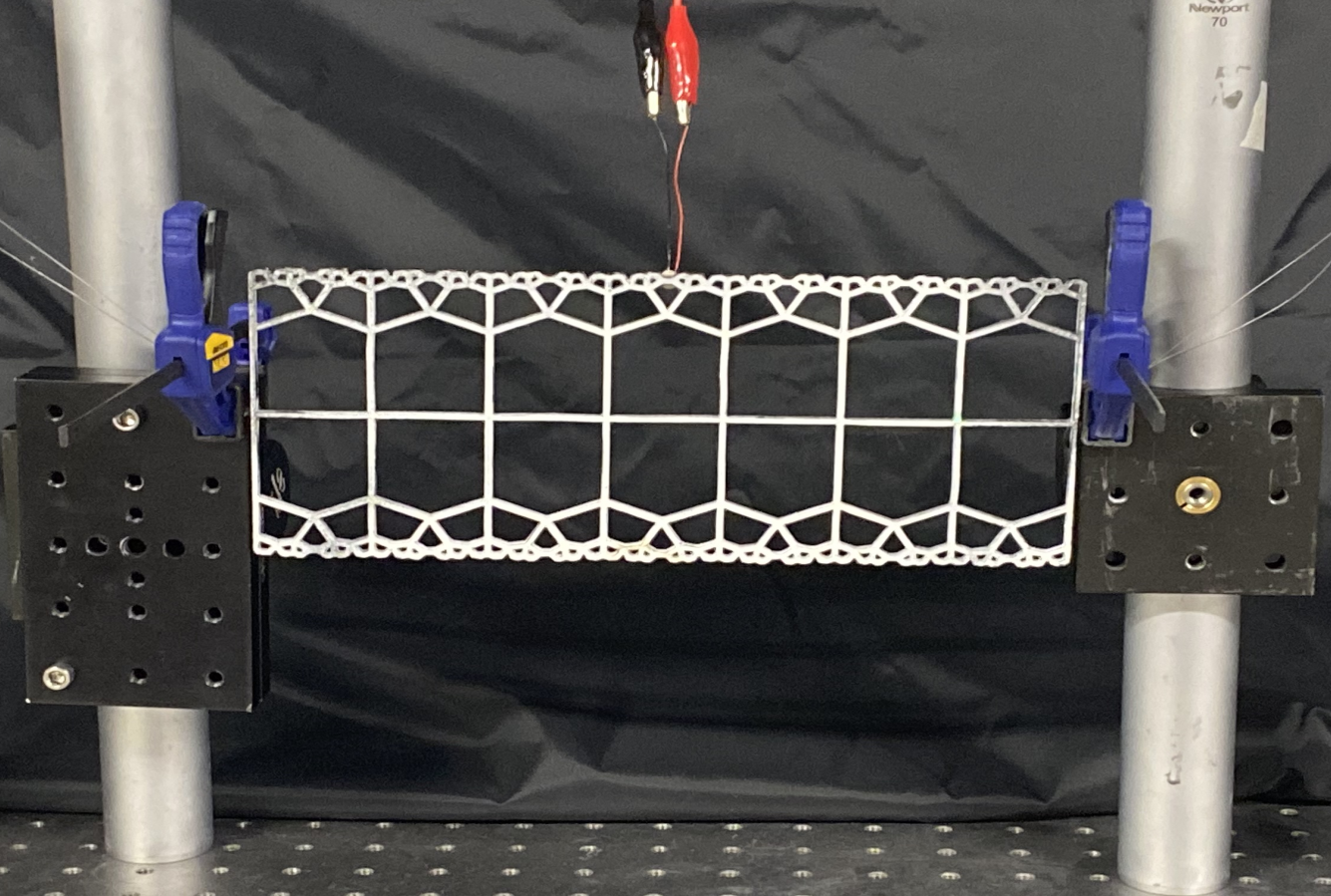}
    \captionsetup{justification=raggedright,singlelinecheck=false}
    \caption{Picture of the measurement setup. The sample was laser cut from an HDPE polymer sheet. A scanning laser vibrometer measured the out-of-plane vibrations induced by a piezoelectric disc here affixed to the center of the top edge. The sample was pinned at the center of its right and left sides by tied nylon string whose ends were clamped to optical posts.}
    \label{Fig5}
\end{figure}

We test the validity of our numerical predictions by conducting dynamic tests on the experimental specimen shown in Fig.~\ref{Fig5}. The specimen is laser cut out of a Proteus HDPE polymer sheet~\cite{proteus} to realize the three-generation \{5,4\} hyperbolic strip lattice. The resulting specimen has beam cross sections that are 2.6 mm wide and 3.8 mm thick. Overall, the lattice is 262.35 mm long and 104.48 mm wide, with seven unit cells along its length. We support the specimen at the center of its right and left outer edges by hitching thin nylon wire to both locations by lark's head knots. These knots are then directly clamped to plates affixed to two optical posts which are mounted onto a vibration-isolated optical table on both sides of the lattice. At these sites, the lattice is free to rotate but restricted in its out-of-plane displacements. In all experiments, the lattice is excited via a 5.0 x 0.4 mm ceramic piezoelectric disc (STEMINC SMD05T04R111WL).

\subsection{Frequency domain experiments}

\begin{figure}
    \centering
    \captionsetup{justification=raggedright,singlelinecheck=false}
    \includegraphics[scale=0.9]{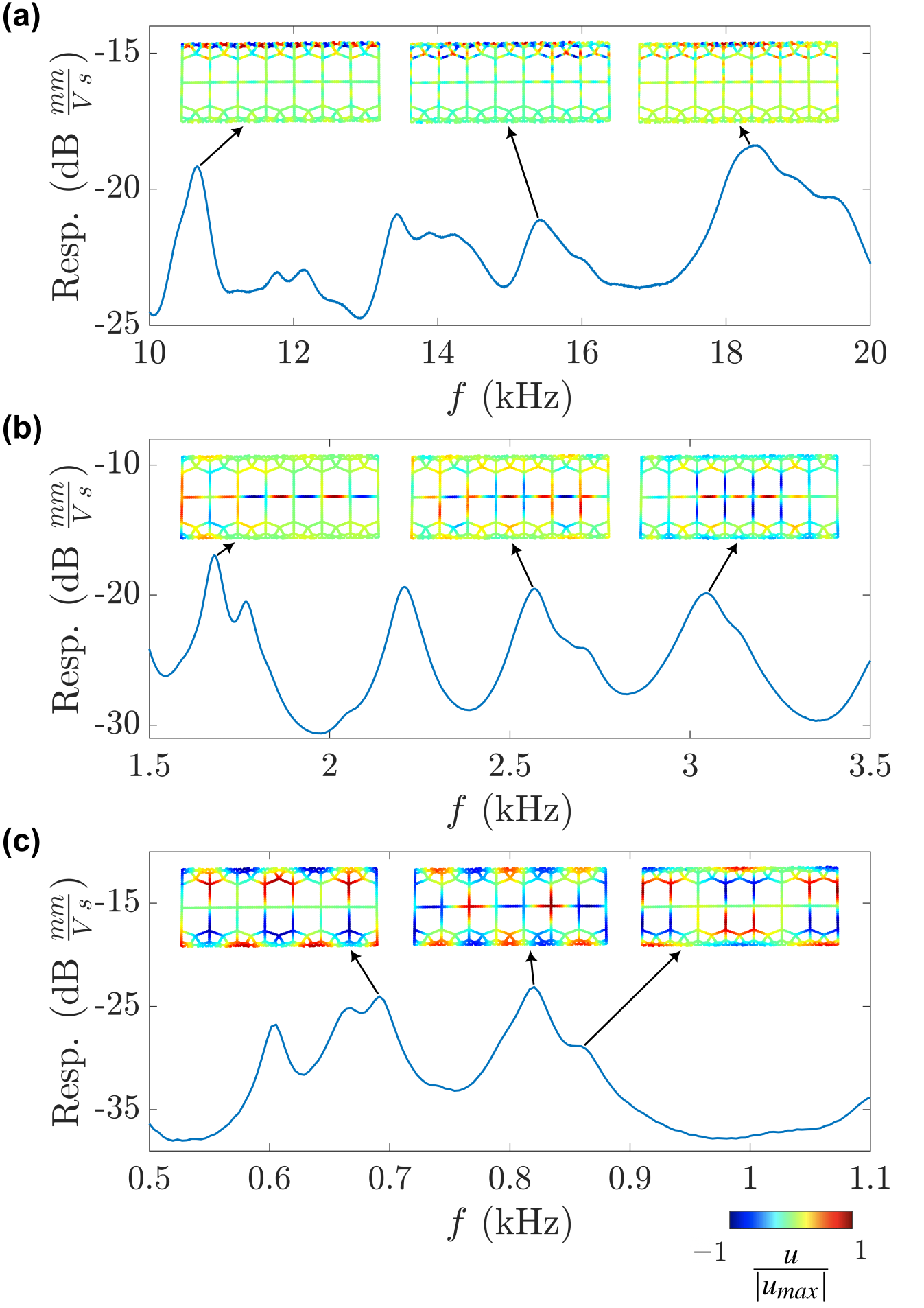}
    \caption{(a), (b), (c) Experimental frequency response in the neighborhood of excitation signals I, II, and III respectively, corresponding to the frequency regions of Fig.~\ref{Fig4}b. Three inset operational deflection shapes in each frequency response plot illustrate the displacement amplitude and relative phase of measured vibration patterns at the corresponding response peaks in the frequency domain. In (a) we observe three boundary shapes with comparable degrees of localization. In (b) we observe three interior shapes with increasing localization with decreasing frequency. In (c) we observe three global shapes.}
    \label{Fig6}
\end{figure}

First, frequency domain tests are conducted to record the frequency response function of the lattice and locate its vibrational modes through the detection of resonances in spatially averaged measurements. We first excite the lattice at the center of its top boundary, as shown in Fig.~\ref{Fig5}. In a second experiment, we excite the lattice at its centroid. In both experiments, we employ a pseudo-random noise input signal with broadband frequency content spanning 0-33 kHz, the range of the numerical states studied in this paper. The velocity response of the lattice is measured by a scanning laser Doppler vibrometer at an evenly distributed set of points sampling the lattice. We obtain the frequency response as the transfer function from the broadband input to the measured velocity response in the frequency domain, which is plotted on a decibel scale over the three frequency ranges of interest in Fig.~\ref{Fig6}a-c. Figure~\ref{Fig6}a,c show portions of the response resulting from the first experiment, where the broadband input is applied at the boundary. These portions are in the spectral neighborhood of the signals denoted as I and III in the numerical studies (see Fig.~\ref{Fig4}). Similarly, Fig.~\ref{Fig6}b shows a portion of the response obtained in the second experiment, where the lattice is excited in its interior. This portion corresponds to the neighborhood of signal II. Inset in each of these figure panels are three examples of measured operational deflection shapes corresponding to resonant peaks. As predicted from numerical simulations, Fig.~\ref{Fig6}a shows boundary modes, Fig.~\ref{Fig6}b shows interior modes, and Fig.~\ref{Fig6}c shows global modes. 

\subsection{Time domain experiments}

\begin{figure*}[!htbp]
    \centering
    \captionsetup{justification=raggedright,singlelinecheck=false}
    \includegraphics{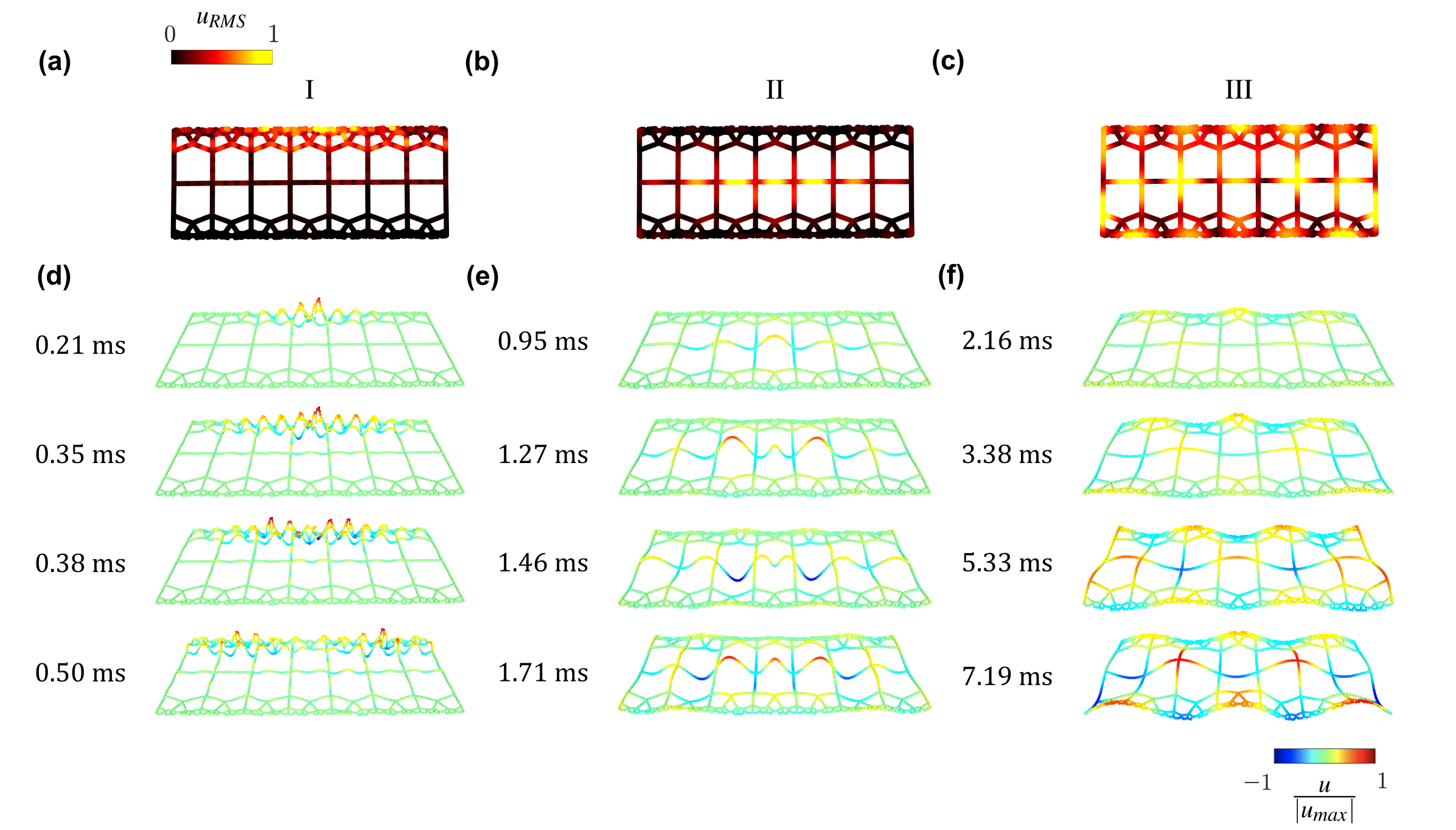} \captionsetup{justification=raggedright,singlelinecheck=false}    
    \caption{(a), (b), (c) Experimental RMS displacement field of the lattice to transient signals I, II, and III respectively. Signals I and III are excited at the center of the top boundary while signal II is excited at the centroid. (d), (e), (f) Snapshots of the experimental time domain response of the hyperbolic strip lattice to signals I, II, and III respectively. Displacements are normalized by the maximum displacement in each considered time instant.}
    \label{Fig7}
\end{figure*}
Next, we measure the response in the time domain for three separate experiments using excitation signals I, II, and III. As functions of time, these signals take the form of Hanning-modulated sinusoids at the central frequencies marked in Fig.~\ref{Fig3}b. As in the frequency response experiments, signals I and III are excited through a piezoelectric disc attached to the center of the top boundary of the lattice, while for signal II, the disc is attached to the lattice centroid. We measure the response in time over the course of three times the duration of each signal (defined as the number of signal oscillations times the period of the signal central frequency). We then take the RMS displacement field in the lattice over the course of this excitation, which produces the plots in Fig.~\ref{Fig7}a,b,c. We observe that these recorded responses are all consistent with the numerical predictions of Fig.~\ref{Fig4}c.

Snapshots in time for each of these three experiments are provided in Fig.~\ref{Fig7}d,e,f. In Fig.~\ref{Fig7}d, we see that over the course of the excitation, waves generated by signal I are confined along the boundary the signal is applied to and exhibit extremely low transmission to the opposite boundary. Figure~\ref{Fig7}e shows the case of an interior response to signal II where elastic waves propagate along the interior region of the lattice, with little motion on the surrounding boundaries. Lastly, Fig.~\ref{Fig7}f shows snapshots of the lattice response to signal III which excites a global behavior in the lattice with high transmission from the point of incidence on the top boundary to the rest of the domain. The experimental results presented here show a strong agreement with numerical predictions and verify the elastic hyperbolic strip's ability to confine propagating waves along particular regions of the lattice based on their frequency content. Such a feature can be leveraged in applications where vibration isolation is desired (see Supplemental Note IV).

\section{Conclusion}
In this paper, we investigate the localized spectral properties of an elastic hyperbolic strip lattice, a lattice which densifies towards its boundaries. We begin by describing the map which takes sites of a circular hyperbolic lattice to the strip domain. We then generate an elastic hyperbolic strip lattice by coupling these sites with structural beams. The spectrum of this lattice structure is numerically computed and its integrated density of states estimated, which reveals the existence of three eigenmode classes: boundary, interior, and global. Example mode shapes of each class are provided. Next, we introduce a relative density of states in order to identify regions of the spectrum which are predominantly populated by boundary, interior, or global modes. Informed by the relative density of states plot, boundary-dominated, interior-dominated, and global-dominated regions of the spectrum are numerically investigated in the time domain. These simulations confirm the boundary-localized, interior-localized, and global response of the lattice to transient inputs centered at different frequencies. Finally, we experimentally confirm the numerical predictions through dynamic testing of a laser-cut specimen whose frequency and time domain responses are measured via laser Doppler vibrometry. Peaks in the experimental frequency response reveal that the relative density of states accurately captures spectral regions of localized or globally-dominated states. Time-averaged displacements and snapshots in time corroborate the numerical findings. The results presented in this work demonstrate the to-date unexplored dynamic properties of a hyperbolic lattice with the capability to confine incident vibrations to localized structural regions along its width. This feature may find applications in waveguiding as well as protection of portions of the structure from the transmission of incident vibrations.

\section{Acknowledgments}
This work was supported by NSF Award No. 2131758.
The authors declare that they have no competing interests.

\bibliography{bibliography}

\newpage
\setcounter{figure}{0}


\renewcommand{\thefigure}{S\arabic{figure}}
\newenvironment{suppfigure}{
    \renewcommand{\figurename}{FIG. 
\renewcommand{\thefigure}{S\arabic{figure}}}
    \begin{figure}
}{
    \end{figure}
}

\newenvironment{Bigsuppfigure}{
    \renewcommand{\figurename}{FIG. 
\renewcommand{\thefigure}{S\arabic{figure}}}
    \begin{figure*}
}{
    \end{figure*}
}



\begin{center}
    {\fontsize{12}{12}\selectfont \textbf{Supplemental Material: Elastic Hyperbolic Strip Lattices}\par}
    \vspace{0.15cm}
\end{center}

\section*{Supplemental Note I: A generalized conformal mapping}

In the main text, we provide a conformal map (Eq. (1)) taking the disk to the strip aligned with the real axis. Here, we generalize this map to allow for rotation, bending, and splitting of the strip along arbitrary axes, all while preserving the lattice network topology. Such a mapping broadens the strip lattice design space and may allow for the creation of path-varying waveguide designs.

\begin{suppfigure}[h]
    \centering
    \includegraphics{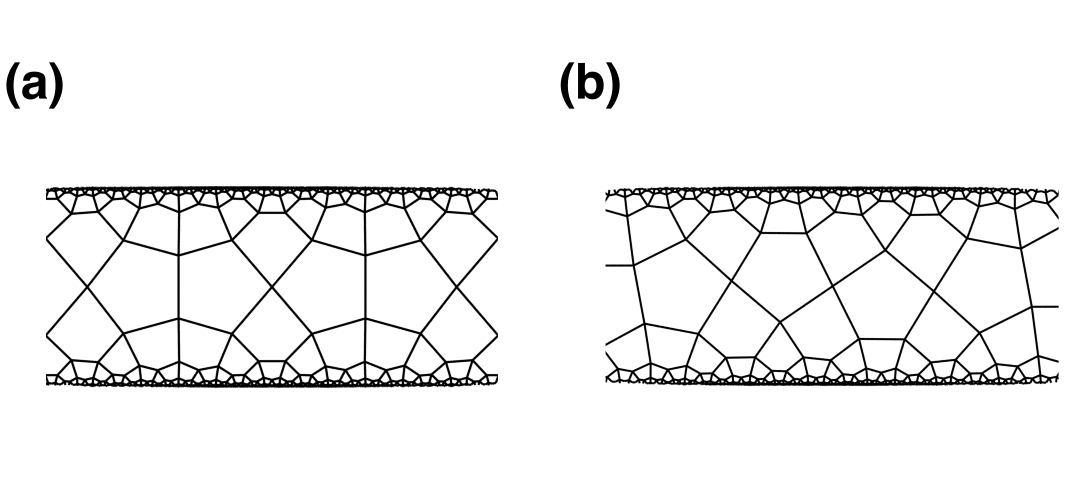}
    \captionsetup{justification=raggedright,singlelinecheck=false}
    \caption{The conformal mapping of two rotated \{5,4\} circular hyperbolic lattices to strip domains. The rotation angles are (a) $\phi=\pi/4$ and (b) $\phi=\pi/6$. The sides are truncated to fit finite domains.}
    \label{rotatedStrips}
\end{suppfigure}

To arrive at the general map, we first compose Eq. (1) of the main text with the pure rotation $h(z) = az$, resulting in

\begin{equation}\tag{SE1}
    (w \circ h)(z)= \frac{4}{\pi} \arctanh{(az)},
    \label{firstRotation}
\end{equation}

\noindent where $a=e^{i\phi}$ is a phase shift rotating points $z$ in the complex unit disk by angle $\phi$. Thus, $z=\pm e^{-i\phi}$ in the pre-rotated disk are made coincident with the conformal map singularities, resulting in a different beam configuration than that in the main text when applied to the same circular \{5,4\} lattice. By varying the rotation angle, a plethora of geometries is possible.

Figure~\ref{rotatedStrips} provides examples of Eq.~\eqref{firstRotation} applied to a circular \{5,4\} lattice with $\phi=\pi/4$ and $\phi=\pi/6$. We see that when $\phi=n\pi/q$ for $n \in \mathbb{Z}$ applied to a \{p,q\} tiling with an origin-centered vertex (and a real axis symmetry when $\phi=0$), we can create a periodic strip lattice. The periodicity converges in the limit of infinite lattice generations, or with sufficiently high generations ($g>10$) one can repeat the central unit cell, as done in the main text.

Next, we take the logarithmic identity 

\begin{align*}\tag{SE2}
       \frac{4}{\pi} \arctanh{(az)} = \frac{4}{\pi}[&\frac{1}{2}\ln(1+az) \\
       &- \ln(1-az)],
\end{align*}

\noindent and apply a second rotation with equivalent phase shift $a$, giving us

\begin{equation}\tag{SE3}
    (h \circ w \circ h)(z)=  \frac{2a}{\pi}[\ln(1+az) - \ln(1-az)].
    \label{fullComposition}
\end{equation}

By Eq.~\eqref{fullComposition}, we observe the composition of a pure rotation, a conformal map, and another pure rotation. The final rotation is superfluous as it is only a rigid body transformation, but it is crucial for the generalization to non-antipodal singularities, which allow for the bending and splitting of the strip. To obtain this general mapping, it helps to define the change of variables $a_1=-a$ and $a_2=a$. Upon this substitution into Eq.~\eqref{fullComposition} and a bit of algebra, we arrive at

\begin{align*}
    (h \circ w \circ h)(z)= -\frac{2}{\pi}[&a_1\ln(1-a_1z) \tag{SE4} \label{general2ptmap} \\ 
    &+a_2\ln(1-a_2z)].
\end{align*} 

$a_1\neq-a_2$ is the general case of non-antipodal singularities. In such a case, we observe the bending of what is otherwise a straight strip. An example of a bent mapping is shown in Fig.~\ref{bentSplit}a.

Recognizing that the complex conjugates of $a_1$ and $a_2$ are the mapping singularities and that Eq.~\eqref{general2ptmap} is the start of a series, we can continue to add analogous logarithmic terms to introduce additional singularities and split the strip along various paths. This map is given by the series

\begin{equation}\tag{SE5}
\tilde{w}(z)=-\frac{2}{\pi} \sum_{j=1}^{N} (b_j \log{(1-a_j z)}),
    \label{splitBand}
\end{equation}

\noindent where $\tilde{w}$ denotes a generalized version of Eq. (1) in the main text, and $N$ is the number of singularities and hence splits. Here we must choose the correct $b_j$ coefficients based on the branch cuts of the added logarithmic terms, as they are generally not equal to $a_j$ as in Eq.~\eqref{general2ptmap}. In Fig.~\ref{bentSplit}b we see an example where $a_j=e^{i((j-1)\pi/2)}$ $b_j=(-1)^ja_j$ for $N=4$ splitting the strip into a cross-like pattern.

\begin{suppfigure}
    \centering
    \includegraphics{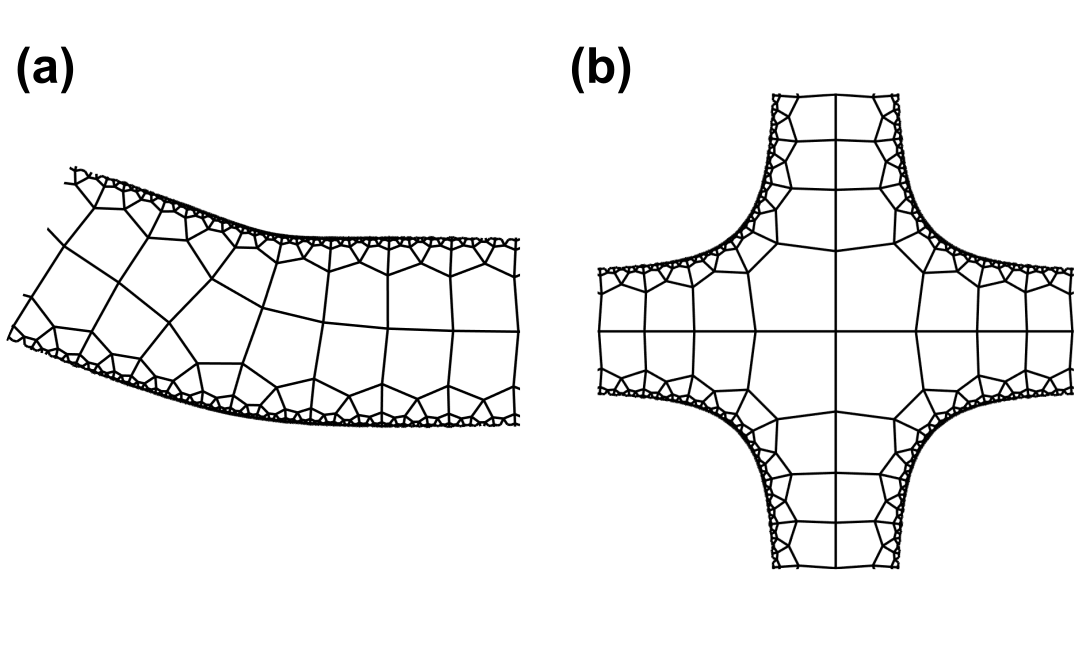}
    \captionsetup{justification=raggedright,singlelinecheck=false}
    \caption{(a) Bent elastic hyperbolic strip lattice obtained by choosing non-antipodal singularities in the conformal map Eq.~\eqref{general2ptmap}. In this case, we select $a_1=1$ and $a_2=e^{i(8\pi/9)}$. (b) Split elastic hyperbolic strip lattice with $a_j=e^{i((j-1)\pi/2)}$ $b_j=(-1)^ja_j$ for $N=4$ in the conformal map of ~\eqref{splitBand}.}
    \label{bentSplit}
\end{suppfigure}

This presented note provides additional information on the novelty of the hyperbolic strip platform for elastic lattices by introducing a series of additional geometries emerging from a generalization of the implemented conformal map. In this paper, we do not explore the dynamics of lattices obtained from these general maps, although we anticipate that the localized properties reported in the main text persist in these rotated, bent, and split geometries. The advantage of these mappings is their modularity. Various bent and split components, when scaled properly, may be joined to create a path-varying lattice domain which may continue to guide localized waves at the correct frequencies. The $\phi$ parametrization is another useful knob that we anticipate can be leveraged for selective wave transport if varied smoothly. The plethora of maps resulting from this general formulation creates a rich playground for elastic hyperbolic lattices yet to be explored using the groundwork for spectral characterization provided in the main text.

\section*{Supplemental Note II: Beam mass and stiffness distributions}

To illustrate the dependence of beam stiffness and mass on generation, this note details the distributions of both quantities for the numerical \{5,4\} strip lattice studied in the paper. The results are given as functions of beam length and sorted by generation. 

The effective stiffness for out-of-plane structural deformation of lattice beam $n$ is formulated by Timoshenko theory as~\cite{thomas1973timoshenko}

  \begin{equation}\tag{SE6}
        k_n = \frac{\beta EI}{L_n^3(1+\frac{12EI}{G\kappa AL_n^2})},
            \label{beam stiffness}
    \end{equation}
    
\noindent where $G$ is the shear modulus, $\kappa =10(1+\nu)/(12+11\nu)$ is the assigned rectangular cross section shear correction factor, and $L_n$ is the length of lattice beam $n$. $\beta$ is a non-dimensional constant arising from the boundary conditions of each beam.

\begin{suppfigure}[h!]
    \centering
    \includegraphics{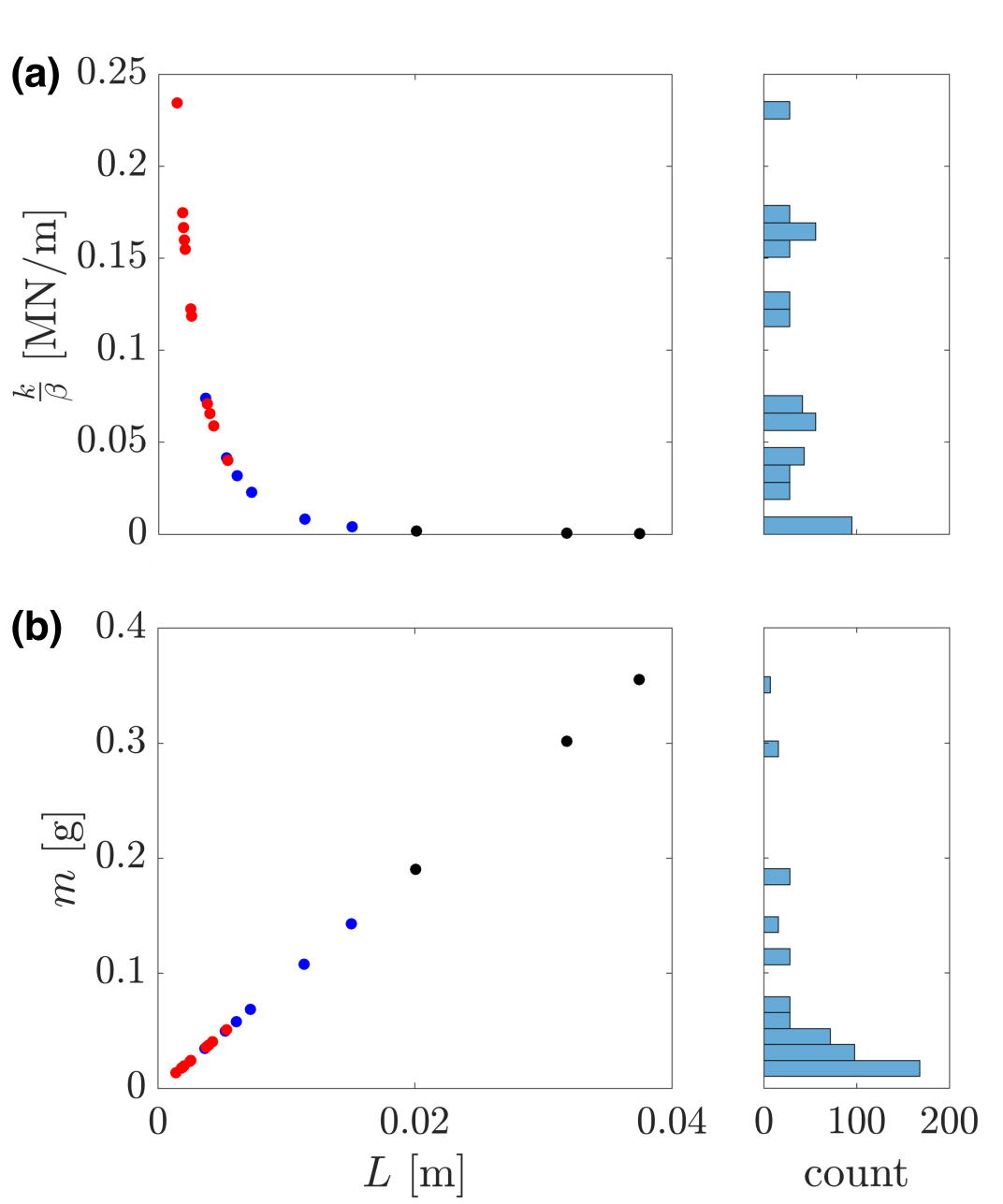}
    \captionsetup{justification=raggedright,singlelinecheck=false}
    \caption{(a) Effective beam stiffness distribution as a function of beam length with an accompanying histogram highlighting stiffness multiplicities. Stiffnesses are normalized by constant $\beta$. (b) Beam mass distribution as a function of beam length, with an accompanying histogram highlighting mass multiplicities. In both panels (a) and (b), black, blue, and red dots correspond to generations 1, 2, and 3 respectively, with beams on the interface of generations assigned to the lower of the two.}
    \label{SuppFig3}
\end{suppfigure}

The mass of beam $n$ is given by

    \begin{equation}\tag{SE7}
        m_n = \rho A L_n.
    \label{beam mass}
    \end{equation}

  \begin{Bigsuppfigure}[!]
        \centering
        \includegraphics{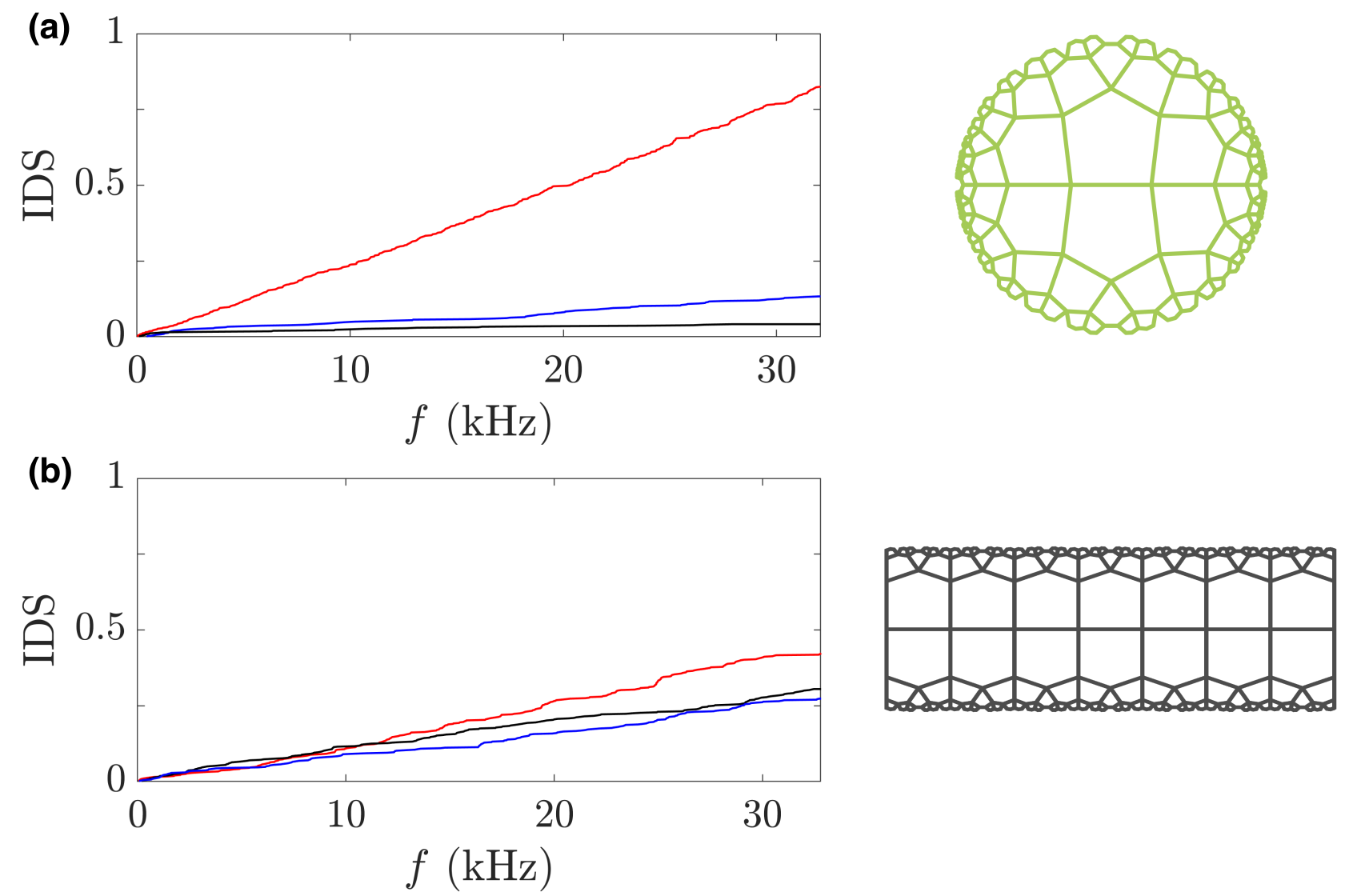}
        \captionsetup{justification=raggedright,singlelinecheck=false}
        \caption{(a) IDS of the \{5,4\} Poincar\'e disk hyperbolic lattice which maps to the finite hyperbolic strip lattice studied in the main text. (b) IDS of the studied \{5,4\} strip lattice (same as Fig. 2a). Geometries corresponding to each IDS are provided to the right of the respective plot (not to scale: the disk lattice radius equals half the strip lattice vertical width). Red, black, and blue curves correspond to boundary, interior, and global mode IDSs.}
        \label{SuppFig4}
\end{Bigsuppfigure}

    After solving Eqs.~\eqref{beam stiffness}, and~\eqref{beam mass} for each beam in the lattice, we plot their distributions as functions of $L_n$ and color each point according to the generation it belongs to. The plotted stiffness is normalized by $\beta$. Figure \ref{SuppFig3} shows these distributions in the form of scatter plots accompanied by histograms that highlight the multiplicities not readily apparent from the scatter plots.

     We see that as we grow the lattice to higher generations, we adjoin higher stiffness, lower mass beams. Though there is some overlap in beam stiffnesses and masses across generations, the mean stiffness and mass per generation is well-separated. This translates to an analogous separation of fundamental frequency scales per generation, and hence different spectral regions dominated by modes in the interior (predominantly $g=$ 1) and boundary ($g=$ 2,\,3). Referring to Fig. 3b of the main text, we observe an initial dominance of modes with significant displacement in generation 1 (interior) which then evolves into a dominance in modes of generations 2 and 3 (boundary). This result is elucidated by Fig.~\ref{SuppFig3} which shows that generation 1 has a lower stiffness to mass ratio than generations 2 and 3, and therefore lower corresponding natural frequencies.

    \section*{Supplemental Note III: IDOS of \{5,4\} Poincar\'e disk lattice}

    In this section, we compare the integrated density of states (IDS) of the \{5,4\} strip lattice provided in the main text to its pre-image, the circular \{5,4\} Poincar\'e disk lattice with the same network topology. The latter is obtained via a finite element eigenfrequency study using the same material properties and beam cross section as the strip lattice in the main text. Figure~\ref{SuppFig4}a and~\ref{SuppFig4}b provide the Poincar\'e lattice IDS and strip lattice IDS respectively for the first 600 modes of each system. Figure~\ref{SuppFig4}b is the same as Fig. 2a, only with a wider plotted range. In plotting these curves, we adopt the same IDS definition as in Eq. (4) of the main text. The localization index used to separate the IDS of the Poincar\'e lattice is identical to that employed in~\cite{patino2024hyperbolic}. Both panels are accompanied by their corresponding system geometry. The radius of the disk (not drawn to scale) equals half the width (short dimension) of the strip which ensures comparable length, and hence frequency, scales. Since the Poincar\'e lattice preserves the network topology of the strip lattice, the Poincar\'e lattice holds a few tiles from higher disk generations on its right and left sides. This is purely an artifact of the mapping, and is required in order to obtain the three-generation strip lattice.

    We observe that the disk lattice IDS is reflective of the general results in~\cite{patino2024hyperbolic}, namely there is a high integrated density of boundary modes (82.5\%). In contrast to~\cite{patino2024hyperbolic}, we here differentiate between global and interior modes of the disk lattice by introducing an interval surrounding the disk localization index threshold $\mathcal{L}_g \in $ [0.45,0.55] in which we classify global modes in the disk lattice. Comparing both Fig.~\ref{SuppFig4}a and~\ref{SuppFig4}b, we observe that boundary modes make up the majority of each spectrum, only that the strip is characterized by an increase in interior localized modes (+26.8 percentage points) and global modes (+13.3 percentage points). In both cases, localized modes (boundary and interior, combined) significantly outnumber global modes, though boundary modes drastically dominate in the disk lattice whereas a closer balance of boundary and interior modes is seen in the strip.

\section*{Supplemental Note IV: Edge to edge transmission}

     \begin{Bigsuppfigure}[t!]
    \centering
    \includegraphics{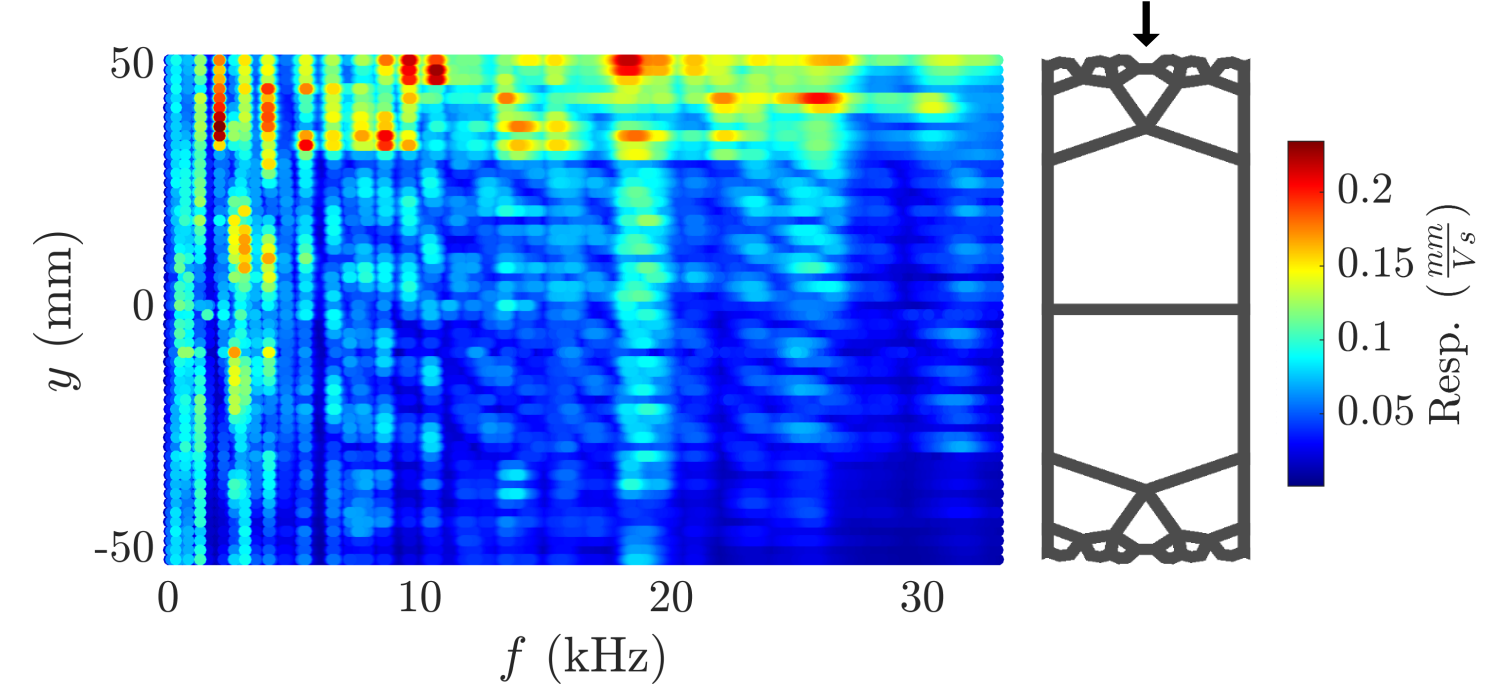}
    \captionsetup{justification=raggedright,singlelinecheck=false}
    \caption{Experimental frequency response to a uniform broadband input signal applied to the center of the top boundary of the elastic hyperbolic strip lattice. The response of the full lattice is plotted as a function of the $y$-position, which evenly samples the experimental lattice 54 times in $y$. A unit cell (right) is provided to illustrate the corresponding $y$-positions of the plotted data, which is for the full strip lattice domain. An arrow indicates the point of the applied dynamic load in the central unit cell of the lattice. The colorbar gives the measured velocity response magnitude.}
    \label{SuppFig5}
\end{Bigsuppfigure}

In this note, we provide supporting results on the wave confinement capabilities of the \{5,4\} elastic hyperbolic strip lattice to validate its potential for vibration isolation applications. Specifically, we provide the experimental frequency response function to a uniform, broadband 0-33kHz chirped input applied to the center of the lattice's top boundary. The results are obtained via laser Doppler vibrometry sampling the lattice at the same set of scan points used in Section Va. This measurement is given as a function of the y-position in the lattice. Along the y-dimension (short dimension/width), the lattice is evenly discretized into 54 sections, resulting in 54 slices of laser scan points. We take the average response in each $y$-slice and plot it sequentially for each measured frequency in Fig.~\ref{SuppFig5}, where the magnitude of the response is given by the colorbar. A unit cell accompanies the figure as a reference for corresponding $y$-positions of the plotted response measurements, although these responses are for slices covering the entire length of the lattice. An arrow points to the position of the applied load in the central unit cell of the lattice.

Figure~\ref{SuppFig5} reveals that the elastic hyperbolic strip lattice attenuates the vast majority of boundary-incident frequencies, preventing them from transmitting through the bulk to the opposite boundary. We see that for $y>$ 30 mm, in the upper second and third generations, the response of the lattice is relatively high for virtually all frequencies, with strong resonances sparsely scattered up to 27 kHz. For $y<$ 30 mm, we observe the nearly immediate decay in the average response, indicating that most of the vibrational energy is localized to the incident boundary. By the bottom of the lattice, near $y<-$50 mm, we observe an extreme attenuation of all frequencies above 3 kHz. Below 3 kHz, we observe one resonance, 2.05 kHz, which is attenuated almost entirely, though below this frequency, there is minimal attenuation. From 2.05 kHz to 5 kHz we observe that the lattice does not attenuate waves smoothly over its length, but rather amplifies its response in its interior before ultimately attenuating its response at the bottom edge. These observations are consistent with the relative density of states plot (Fig. 3b) in the main text, which indicates global behavior below 2 kHz and interior-mode behavior between 2-5 kHz, above which the boundary modes of the system dominate on average until 28 kHz.

The results in Fig.~\ref{SuppFig5} demonstrate that the strip lattice can be used in vibration isolation applications as it successfully attenuates a broad range of frequencies from transporting across its width. These results are supported by the findings in the main text.


\end{document}